\documentclass[twoside]{report}
\usepackage{iwsm}
\usepackage{graphicx}
\usepackage{mathrsfs}
\usepackage{amsmath, amssymb}
\begin{document}

%***********************************************************************
% PLEASE INSERT YOUR CONTENT FROM HERE
%***********************************************************************

% Title and running title to be used as left header:
\title{Bayesian shape modelling of cross-sectional geological data}
\titlerunning{Bayesian shape modelling}

% Authors and running list of authors to be used as right header:
\author{Thomai Tsiftsi\inst{1}, Ian H. Jermyn\inst{1}, Jochen Einbeck\inst{1}}
\authorrunning{Tsiftsi et al.}    %% use \authorrunning{Surname 1} if only 1 author
                                    %% use \authorrunning{Surname 1 and Surname2} if two authors
                                    %% use \authorrunning{Surname 1 et al.} if more than two authors

% Institutes of all authors
% Include city and country of each institute, do not include the full address.
\institute{University of Durham, Department of Mathematical Sciences, Statistics and Probability Group, UK}

% E-mail of presenting author for correspondence
\email{thomai.tsiftsi@durham.ac.uk}

% Brief abstract of the paper:
% \abstract{The recognition of objects in images is an important problem in many branches of science such as medical image analysis, military target recognition and geological classification. A way to address this problem is to represent these objects by using certain features such as their shape. Using statistical models, one can classify these shapes into their retrospective categories. We apply statistical shape analysis to understand the shape and variability of geological sandbodies.}

\abstract{Shape information is of great importance in many applications. For example, the oil-bearing capacity of sand bodies, the subterranean remnants of ancient rivers, is related to their cross-sectional shapes. The analysis of these shapes is therefore of some interest, but current classifications are simplistic and ad hoc. In this paper, we describe the first steps towards a coherent statistical analysis of these shapes by deriving the integrated likelihood for data shapes given class parameters. The result is of interest beyond this particular application.} 

% Keywords (at most 5):

% Produce the title:
\maketitle

%***********************************************************************

% Sections and subsections (do not use lower levels):

\section{Introduction}
\noindent
Sand bodies, the sedimentary, subterranean remnants of ancient rivers, are important to both geology and the petroleum industry. In particular, their cross-sectional shapes help determine their oil-bearing capacity. Current classification schemes for sand body shapes are qualitative, simple, and ad hoc, and so there is a need for a quantitative analysis with the help of statistical models. There are several problems of interest: estimation of shape class parameters given labelled data shapes (a `data shape' is an ordered set of points in $\mathbb{R}^{2}$); classification of new data shapes; and unsupervised classification. Parameter estimation is described by the probability $\mathbb{P}(w|y, c)$, where $w$ denotes the shape class parameters and $y$ the dataset, which consists of several data shapes, together with their class labels $c$. By Bayes' theorem, this is given by: 
    \begin{equation}
        \mathbb{P}(w|y, c)\propto  
        \mathbb{P}(y|w, c)\mathbb{P}(w) \:.
    \end{equation}

\noindent
In this, as in all of the above problems, the major task is to calculate the likelihood $\mathbb{P}(y|w, c)$. This is the problem addressed in this paper. The problem is not unique to the sand body application: it occurs in many applications of shape modelling, and is thus of broad interest.

\section{The likelihood}
The calculation of the likelihood is complicated due to the presence of many nuisance parameters that must be integrated over. The partitioned likelihood is: 
    \begin{multline}
        \mathbb{P}(y|w, c) = \\
        \sum_{b \in B}
        \int ds~ dg~ d\beta ~d\sigma~ 
        \mathbb{P}(y|\sigma, b, s, g, \beta)
        \mathbb{P}(\sigma) 
        \mathbb{P}(b) 
        \mathbb{P}(s|\beta)
        \mathbb{P}(g)  
        \mathbb{P}(\beta|w, c) \:,
        \label{eq:likelihood}
    \end{multline}

where we have made a number of simplifying independence assumptions. Here  $\mathbb{P}(y|\sigma, b, s, g, \beta) = \exp{\left(-\frac{1}{2\sigma^2}\sum_{i=1}^{N} |y(i)-g\circ \beta (s(b^{-1}(i)))|^2 \right)}$ models errors in shape point collection as Gaussian white noise.

\medskip
\noindent
In the above expression, $\beta$ is the underlying sand body shape modulo similarity transformations, which comes from a class $C$ with parameters $w$, while $g \in G \equiv SO(2)\ltimes \mathbf{R}^2 \times \mathbf{R}_+$ is a similarity transformation generating the full sand body shape $g\beta$. Data formation is modelled as a sampling $s$ of $N$ points around the sand body shape, and a bijection $B \ni b: [1,...n ]\rightarrow [1,...n] $ relating each point of the sand body shape to a unique point of the data shape, giving $g\beta(s(b^{-1}))$, plus the above Gaussian noise with variance $\sigma$.

\medskip
\noindent
In previous work, \emph{e.g.}\ Dryden and Mardia (1998) or Srivastava and Jermyn (2009), an algorithmic approach was taken to the integrals over the group $G$, using the Procrustes algorithm to compute a zeroth order Laplace approximation. Here we carry out the group integrations, and the integration over $\sigma$, analytically, resulting in a closed form expression. This is the main contribution of this paper.

\medskip
First, we have to choose the priors for $g$ and $\sigma$. Jeffrey's joint prior for $g$ and $\sigma$ was calculated to be $\frac{a~\text{Var}(\beta(s))}{\sigma^5}$, but this leads to a non-normalizable likelihood. Instead, a regularized version was employed. A Gaussian prior was used for translations; a uniform prior for rotation angle; and a Rayleigh prior for scaling. These choices break translation invariance by effectively limiting the size of the two-dimensional domain in $\mathbb{R}^{2}$ in which the shape points lie, and break scale invariance by effectively limiting the range of scales considered. With these priors, the result of the integrations over translations, rotations, and scalings is: 
    \begin{multline}
        \mathbb{P}(y|b, s, \beta) = \\
        \frac{1}{Z}\frac{1}{\sigma^{2n}}
        \exp
        \left\{
            \frac{-1}{2\sigma^2}
            \left(
                \tilde{n}\tilde{\text{Var}(y)}
                -\frac{\tilde{n}^2\tilde{\text{Cov}^2}(\beta                 (s(b^{-1})),y)}{\tilde{n}\tilde{\text{Var}}                 (\beta (s(b^{-1})) + 1/B^{2}}
            \right)
        \right\} \:,
    \end{multline}

\noindent
where:
\begin{itemize}
\item   

$\tilde{n}=n+\frac{1}{D^2}$

\item 

$\tilde{\text{Var}}(y)=\frac{1}{\tilde{n}}\left[\sum_{i}|y_{i}|^2-\frac{1}{\tilde{n}}\sum_{i}\sum_{j}y_{i}y_{j}  \right]$

\item 

$\tilde{\text{Cov}}(\beta (s(b^{-1})),y)=\frac{1}{\tilde{n}}\left[ \sum_{i}\beta (s(b^{-1}(i)))y_{i}-\frac{1}{\tilde{n}}\sum_{i}\sum_{j}\beta (s(b^{-1}(i))y_{j}\right]$

\end{itemize}

\medskip
\noindent
which are regularized versions of the number of points and the variance. $B, D, \alpha, c$ are appropriate regulators and $Z$ the normalization constant. 

\medskip
A $\Gamma$ prior was used for $\sigma$. After integration, this leads to:
    \begin{multline}
        \mathbb{P}(y|b, s, \beta) = 
        \frac{\Gamma(n+\alpha)}{Z} \times
        \left[  
            \tilde{n}\tilde{\text{Var}(y)}-\frac{\tilde{n}^2             |\tilde{\text{Cov}}(\beta (s(b^{-1})),y)|^{2}}{
            \tilde{n}\tilde{\text{Var}}(\beta) 
            (s(b^{-1})) + 1/B^{2}} + 2c   
        \right]^{n-\alpha} \:.
    \end{multline}

\noindent This expression is the main result of the paper. It applies to any shape modelling application in which white Gaussian noise is added to a discrete set of shape points. 

\medskip
\noindent
To finish the calculation of the likelihood, we have to perform the $s$ and $\beta$ integrations, and the $b$ summation, in equation~\eqref{eq:likelihood}. This we do using Monte Carlo techniques. We use a uniform distribution on $[0, 1]^{N}$ for samplings $s$, while $\beta$, which consists of positive quantities such as aspect ratios and lengths, is modelled with $\Gamma$ distributions, whose parameters over all classes constitute $w$. In previous work, when the group integrations were approximated using the Laplace approximation and the Procrustes algorithm, the sum over bijections could be approximated, again in a zeroth order Laplace estimation, using the Hungarian algorithm, since only a linear assignment problem was involved. In the integrated likelihood derived here, the terms involving $\tilde{\text{Cov}(\beta (s(b^{-1})),y)}$ complicate the situation, and turn the linear assignment problem into quadratic assignment, which is NP-hard. Instead of using the Laplace approximation, we approximate the full summation using Monte Carlo, with a uniform prior on $b$. 

\section{Parameter estimation}
\begin{figure}
    \begin{center}
        \includegraphics[width=0.8\textwidth]{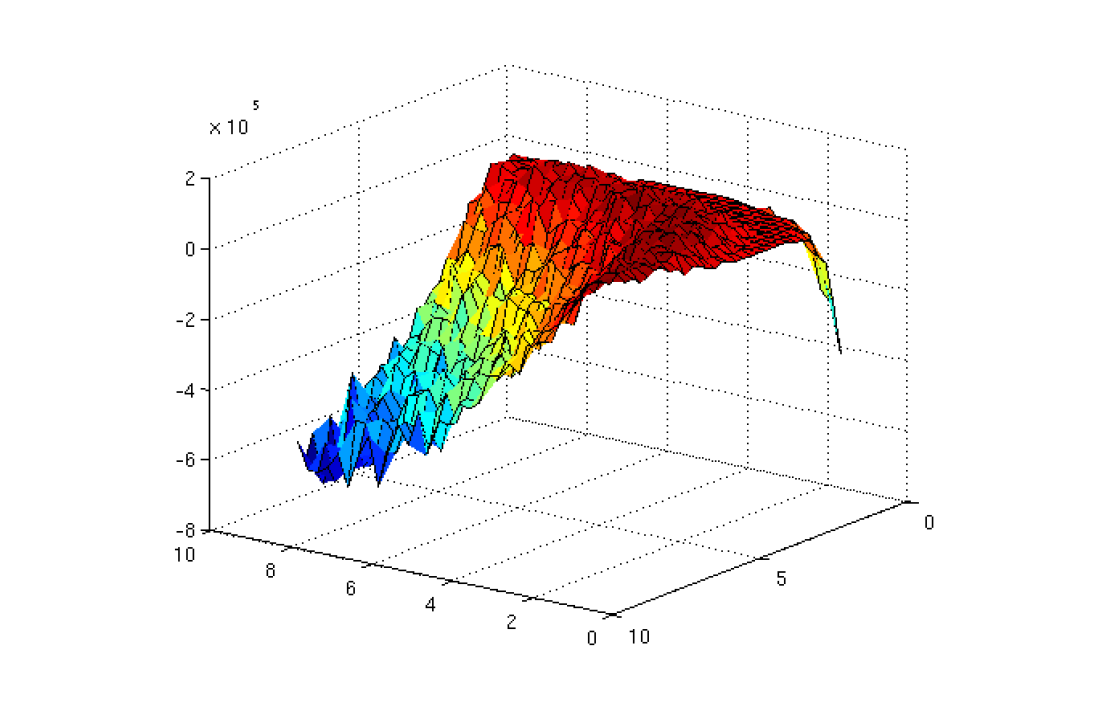}
    \end{center}
    \caption{\label{fig1}The likelihood surface}
\end{figure}

The above result can be used to estimate the class parameters $w$ given data shapes from each class. Figure~\ref{fig1} shows an example of a likelihood surface, in this case computed for the simplest case of a rectangle modulo similarities. This is a one-dimensional shape space that can be parametrized by the aspect ratio. There is thus one Gamma distribution involved, and $w$ is two-dimensional. The surface is rough, since only a coarse grid in parameter space was used to reduce simulation time. 

Maximum likelihood estimation was carried out using a built-in Matlab optimization function. The algorithm correctly navigated towards the maximum of the likelihood surface, but convergence was slow. We hope to improve this in the future by using a different maximization algorithm. 

\section{Conclusion}
The main contribution of this paper is the analytical evaluation of the integrals over the similarity group and the noise variance in a model for shape data. The application considered here is to the classification of the cross-sectional shapes of sand bodies, but the same techniques apply to any shape model involving Gaussian noise. This work is still in progress but shows promising signs of improving the current classification and estimation methods employed by geologists. There are technical obstacles in the form of numerical integrations which we hope to overcome in the near future.

\references
\begin{description}
\item[Dryden, I.L. and Mardia, K] (1998).
    {\it Statistical shape analysis}.
    J. Wiley.
\item[Srivastava, A. and Jermyn, I.H.] (2009)
    Looking for shapes in two-dimensional, cluttered point clouds.
    {\it IEEE Trans. Patt.\ Anal.\ Mach.\ Intell.}, {\bf 31(9)},
    1616\,--\,1629.
\end{description}

\textbf{Notes} 

\medskip
In the manuscript version published in the conference proceedings volume, there was a small error in expression (4). This is now corrected.

\end{document}